\documentclass[]{raa}           

\usepackage{graphicx,times}
\usepackage{natbib}
\usepackage{amssymb,amsmath}
\bibpunct{(}{)}{;}{a}{}{,}

\usepackage[pagebackref=true]{hyperref}

\begin{document}

   \title{Implication of the Stellar Mass Density of High-$z$ Massive Galaxies from JWST on Warm Dark Matter}

 \volnopage{ {\bf 20XX} Vol.\ {\bf X} No. {\bf XX}, 000--000}
   \setcounter{page}{1}

   \author{Hengjie Lin \inst{1,2}
   \and Yan Gong \inst{1,2,3*}
   \and Bin Yue \inst{1}
   \and Xuelei Chen \inst{1,2,4,5}
   }

   \institute{ National Astronomical Observatories, Chinese Academy of Sciences, Beijing 100101, China; {\it gongyan@bao.ac.cn}\\
   \and
     University of Chinese Academy of Sciences, Beijing 100049, China; \\
    \and
     Science Center for China Space Station Telescope, National Astronomical Observatories, Chinese Academy of Sciences, 20A Datun Road, Beijing 100101, China; \\
    \and
      Department of Physics, College of Sciences, Northeastern University, Shenyang 110819, China \\
    \and
      Center for High Energy Physics, Peking University, Beijing 100871, China \\
\vs \no
   {\small Received 20XX Month Day; accepted 20XX Month Day}
}

\abstract{A significant excess of the stellar mass density at high redshift has been discovered from the early data release of James Webb Space Telescope ($\it{JWST}$), and it may require a high star formation efficiency. However, this will lead to large number density of ionizing photons in the epoch of reionization (EoR), so that the reionization history will be changed, which can arise tension with the current EoR observations. Warm dark matter (WDM), via the free streaming effect, can suppress the formation of small-scale structure as well as low-mass galaxies. This provides an effective way to decrease the ionizing photons when considering a large star formation efficiency in high-$z$ massive galaxies without altering the cosmic reionization history. On the other hand, the constraints on the properties of warm dark matter can be derived from the $\it JWST$ observations. In this work, we study WDM as a possible solution to reconcile the $\it JWST$ stellar mass density of high-$z$ massive galaxies and reionization history. We find that, the $\it JWST$ high-$z$ comoving cumulative stellar mass density alone has no significant preference for either CDM or WDM model. But using the observational data of other stellar mass density measurements and reionization history, we obtain that the WDM particle mass with $m_{\text{W}} = 0.51^{+0.22}_{-0.12}$ keV and star formation efficiency parameter $f_{*}^0>0.39$ in 2$\sigma$ confidence level can match both the $\it JWST$ high-$z$ comoving cumulative stellar mass density and the reionization history.
}

   \authorrunning{H. J. Lin et al.}            %author_head in even pages
   \titlerunning{WDM constraint from $\it{JWST}$}  % title_head in odd pages
   \maketitle

%________________________________________________ sections below
% 
\section{Introduction}\label{sec:introduction}

High redshift (high-$z$) galaxies play an important role in the Epoch of Reionization (EoR), which is considered to be the last phase transition of the Intergalactic Medium (IGM) in the cosmic history. Studying the high-$z$ galaxies allows us to have a deeper understanding on the formation and evolution of early galaxies and the cosmic reionization history. The launch of the James Webb Space Telescope ($\it{JWST}$) opens a new era to study the high redshift object in 
the Universe \citep{Finkelstein-2022, Finkelstein-2023}. The $\it JWST$ early observational data release provides valuable information about the properties of high-$z$ galaxies, which may challenge the current galaxy formation theory under the $\Lambda$CDM model \citep{Boylan-2023, Mason-2023, Menci-2022, Mirocha-2023, Naidu-2022-a, Naidu-2022-b}.

In particular, \cite{Labbe-2023} may find a serious tension with the standard galaxy formation theory. Using the $\it{JWST}$ Cosmic Evolution Early Release Science (CEERS) sample (detected by the JWST/NIRCam instrument), they identified six galaxies with stellar masses $M_{*} = 10^{10}\sim10^{11} h^{-1} M_{\odot}$ at $7.4 \leqslant z \leqslant 9.1$. The corresponding comoving cumulative stellar mass density is about 20 times higher at $z \sim 8$ and about three orders of magnitude higher at $z \sim 9$ than the prediction from the star formation theory in standard $\Lambda$CDM cosmology, which is based on previous observations \citep{Stark-2013, Song-2016, Bhatawdekar-2019, Kikuchihara-2020, Stefanon-2021}. This huge abundance excess may due to issues of galaxy selection, measurements of galaxy stellar mass and redshift, dust extinction, and sample variance \citep{Endsley-2022, Ferrara-2022, Ziparo-2023, Adams-2023}. But if this result is confirmed by future spectroscopic observation, for example the follow-up observation by the JWST/NIRSpec, it will be an huge challenge to the $\Lambda$CDM model. There are many efforts trying to solve this tension, for example, the primordial non-Gaussianity in the initial conditions of cosmological perturbations \citep{Biagetti-2023}; the rapidly accelerating primordial black holes \citep{Liu-2022, Yuan-2023}; the Fuzzy Dark Matter (FDM) \citep{Gong-2023}; axion miniclusters or primordial black holes \citep{Hutsi-2023}; the cosmic string \citep{Jiao-2023}; a gradual transition in the stellar Initial Mass Function (IMF) \citep{Trinca-2023}; modified $\Lambda$CDM power spectrum \citep{Prarshari-2023, Hamsa-2023}.

A direct way to explain the observation of \cite{Labbe-2023} in theory is enhancing the star formation efficiency $f_{*}$. Some previous observations prefer a low star formation efficiency ($f_{*} < 0.1$). But since $f_{*}$ depends on  complicated astrophysical processes which remain unclear, it is theoretically possible to have a larger $f_{*}$ at high redshifts. On the other hand, although assuming large $f_{*}$ at high redshifts may help to solve or reduce the tension between the current model and the $\it{JWST}$ observation, it may raise new problems. If we have higher star forming efficiency at high-$z$, the total number of ionizing photons in the EoR will significantly increase, and then the EoR may end much earlier which could be in tension with the EoR history observations \citep{WMAP9, Planck18}. A possible way to resolve this problem is to enhance the formation rate of massive galaxies, but suppress the formation of low mass galaxies at the same time. A mass-dependent star formation efficiency $f_{*}(M)$ can suppress the star formation in small halos, but may still not enough to reduce the huge tension.

In \cite{Gong-2023}, they propose to use fuzzy dark matter (FDM) to solve this problem. The FDM is expected to have extremely low mass of $10^{-27} \lesssim m_{a} \lesssim 10^{-19}$ eV, and it can suppress the abundance of low-mass haloes via the  effective quantum pressure arises from its galaxy-size de Broglie wavelength, which can successfully explain both $\it JWST$ and EoR observations. In this work, we study warm dark matter (WDM) as an alternative solution, and explore the implications on the WDM properties with the $\it JWST$ data. In the WDM scenario, WDM particles with mass of a few keV are much lighter than the standard CDM particles, allowing them to remain relativistic for longer in the early universe and to retain a non-negligible thermal velocity dispersion. This thermal velocity dispersion will cause a so-called free streaming effect, making WDM particles escape out of the high density region and resulting in a suppression of the structure growth on small scales \citep{Blumenthal-1982, Bode-2001, de-Vega-2012}. Then the abundance of low-mass haloes as well as galaxies will be significantly suppressed, and this insight us to consider WDM as a solution to explain the $\it{JWST}$ observation \citep{Labbe-2023}.
In this work, we assume a flat universe with $\Omega_{\text{m}} = 0.3153$, $\Omega_{\text{b}} = 0.0493$, $h = 0.6736$, $\sigma_{8} = 0.811$, $n_{\text{s}} = 0.9649$ \citep{Planck18} .

\section{Model}\label{sec:model}

\subsection{WDM Halo Mass Function}\label{subsec:HMF}

In order to analysis the impact on the halo abundance by WDM, we need to calculate the halo mass function. A conventional choice for the halo mass function is obtained from the ellipsoidal collapse model \citep{Press-Schechter, Sheth-Tormen-1999, Sheth-Tormen-2001, Cooray-2002}, which can be expressed as
\begin{equation}
    n(M)dM = \frac{\bar{\rho}}{M} f(\nu) d\nu,
    \label{eq:dndm}
\end{equation}
where $M$ is the halo mass, $\bar{\rho}$ is the mean comoving matter density, and
\begin{equation}
    f(\nu) = A \left[ 1 + (q\nu^2)^{-p} \right]e^{-q\nu^2/2}.
    \label{eq:fv}
\end{equation}
Here $p = 0.3$, $q = 0.707$, $A = 0.2161$ are the parameters derived from simulations, and $\nu(M) = \delta_{\text{c}}/\sigma(M)$. 

In the CDM case, we assume the critical overdensity is independent to the mass scale, and $\delta_{\text{c}}$ is the critical overdensity barrier for collapse, which can be derived analytically and given by $\delta_{\text{c}} \approx 1.686$ for the spherical collapse of CDM. On the other hand, in the WDM case, we expect that the free streaming effect will allow the WDM particle escape out of the collapsing region, and as the result, the collapse of WDM will become more difficult on small scales. \cite{Barkana-2001} studied the collapse thresholds for WDM by performing spherical collapse simulations. They find that, when the halo mass is below a  characteristic mass scale, the threshold for collapse will increase rapidly with decreasing mass. \cite{Benson-2013} provided a fitting function to describe the mass dependence of $\delta_{\text{c}}$ as
\begin{equation}
    \delta_{\text{c}}(M,z) = \mathcal{G}(M) \frac{\delta_{\text{c}}^0}{D(z)},
    \label{eq:delta_c-M-z}
\end{equation}
where
\begin{equation}
\begin{split}
    \mathcal{G}(M) = (1 - h(x))\text{exp}\left[ \frac{0.31687}{\text{exp}(0.809x)} \right] \\
    + h(x) \frac{0.04}{\text{exp}(2.3x)}.
    \label{eq:G-M}
\end{split}
\end{equation}
Here $x = \log (M/M_{\text{J}})$, and $M_{\text{J}}$ is the effective Jeans mass of the WDM as defined by \cite{Barkana-2001}:
\begin{equation}
\begin{split}
    M_{\text{J}} = 3.06\times 10^8 \left( \frac{1+z_{\text{eq}}}{3000} \right)^{1.5} \left( \frac{\Omega_{\text{m}}h^2}{0.15} \right)^{1/2} \\
    \times \left( \frac{g_{\text{W}}}{1.5} \right)^{-1} \left( \frac{m_{\text{W}}}{1\text{ keV}} \right)^{-4} M_{\odot},
    \label{eq:M_J}
\end{split}
\end{equation}
where $z_{\text{eq}} = 3600 \left( \frac{\Omega_{\text{m}}h^2}{0.15} 
\right) - 1$ is the redshift of matter-radiation equality, $g_{\text{W}} = 1.5$ is the effective number of degrees of freedom for the spin$-\frac{1}{2}$ fermion, and $h(x) = 1/[1 + \text{exp}(10x + 24)]$ is an auxiliary fitting function. 
$\sigma(M)$ can be calculated as the variance of the linear matter overdensity field when smoothed on a comoving scale $R$:
\begin{equation}
    \sigma^2(R) = \int_{0}^{\infty} 4\pi \left(\frac{k}{2\pi}\right)^3P^{\text{lin}}(k)W^2(kR)d{\rm \,ln}k,
    \label{eq:sigma-R}
\end{equation}
where $W(x) = (3/x^3)\left(\sin x - x\cos x\right)$ is the Fourier transform of a spherical top-hat filter window function. We can rewrite $\sigma(R)$ in terms of mass by relate the comoving scale $R$ with the the mass enclosed within this scale as $M = (4/3)\pi R^3 \bar{\rho}$. $P^{\text{lin}}(k)$ is the linear matter power spectrum, and it can represent the CDM and WDM linear power spectrum in this study. 

Since WDM can suppress the matter clustering on small scales via the free streaming effect, the WDM linear matter power can be calculated by \citep{Bode-2001}:
\begin{equation}
    P^{\text{lin}}_{\text{WDM}}(k,z) = P^{\text{lin}}_{\text{CDM}}(k,z)T_{\text{W}}^2(k).
    \label{eq:power_spectrum}
\end{equation}
The CDM linear power spectrum $P_{\text{CDM}}(k,z)$ is obtained by CAMB \citep{CAMB}, and the transfer function $T_{\text{W}}(k)$ is assumed to be redshift independent, and it is given by
\begin{equation}
    T_{\text{W}}(k) = [1+(\alpha k)^{2\mu}]^{-5/\mu},
    \label{eq:T_w}
\end{equation}
Where the fitting parameters are $\mu = 1.12$, $\alpha = 0.074(m_{\text{W}}/1\text{ keV})^{-1.15}(0.7/h)$ Mpc. 

\begin{figure}
    \centering
    \includegraphics[scale = 0.5]{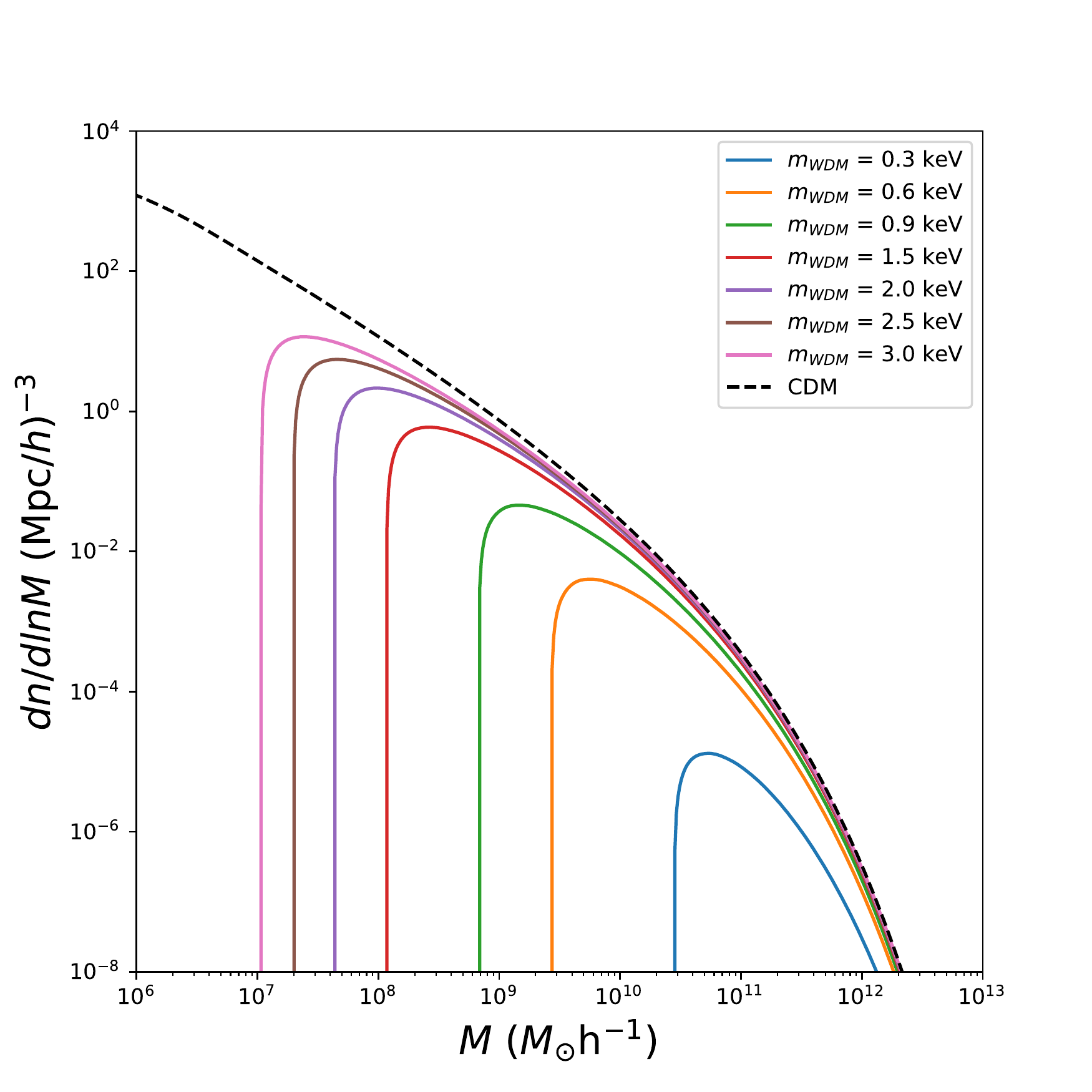}
    \caption{The halo mass functions at redshift $z = 8$ for different WDM masses (solid lines). The CDM case is also shown for comparison (black dash line). We can see that WDM can significantly suppress the formation of low-mass halos, and the suppression mass scale increases with the WDM particle mass decreases.}
    \label{fig:dndm}
\end{figure}

In Fig.\ref{fig:dndm} we show the halo mass functions for both WDM and CDM cases at redshift $z = 8$, and we can find that the abundance of the low-mass halos are strongly suppressed by the free streaming of WDM. So the contribution of ionizing photons from low-mass galaxies will also be suppressed, and we are allowed to have a larger number density for massive galaxies without altering the reionization history. 

\subsection{Stellar Mass Density and Reionization History}\label{subsec:rho_star-and-reion}

After obtaining the halo mass function, the comoving cumulative halo mass density with halo mass greater than $M$ can be estimated by : 
\begin{equation}
    \rho(>M,z) = \int_{M}^{\infty} dM^{'} M^{'} n(M^{'},z).
    \label{eq:rho-M}
\end{equation}
By considering the $M_{*} - M$ relation $M_{*} = \left( \Omega_{\text{b}}/\Omega_{\text{m}}\right) f_{*} M = \epsilon M$, the cumulative stellar mass density with stellar mass larger than $M_{*}$ can be written as :
\begin{equation}
    \rho_{*}(>M_{*},z) = \epsilon \rho(>M_{*}/\epsilon,z).
    \label{eq:rho-star}
\end{equation}
The star formation efficiency $f_{*}$ can be considered as a mass-dependent quantity. In this work, we assume a double power-law form of $f_{*}(M)$ , and following \cite{Mirocha-2017} we have
\begin{equation}
    f_{*}(M) = \frac{2f_{*}^0}{\left( \frac{M}{M_{\text{p}}} \right)^{\alpha_{\text{lo}}} + \left( \frac{M}{M_{\text{p}}} \right)^{\alpha_{\text{hi}}}},
    \label{eq:f_star}
\end{equation}
where $f_{*}^0$ is the star formation efficiency at its peak mass $M_{\text{p}}$, and $\alpha_{\text{lo}}$ and $\alpha_{\text{hi}}$ describe
the power-law index at low and high masses. In our work, we adopt $M_{\text{p}} = 2.8 \times 10^{11} M_{\odot}$, $\alpha_{\text{lo}} = 0.49$ and $\alpha_{\text{hi}} = -0.61$. The parameter $f_{*}^0$ will be adjusted to match the $\it{JWST}$ data in this work.

The star formation efficiency is tightly related to the emission of ionizing photons, and thus can change the cosmic reionization history. Usually, we can explore the reionization history by investigating the redshift evolution of the hydrogen volume filling factor $Q_{\text{HII}}(z)$ by solving the differential equation as below \citep{Madau-1999, Wyithe-2003}:
\begin{equation}
    \frac{dQ_{\text{HII}}}{dt} = f_{\text{esc}} \frac{\dot{n}_{ion}}{\bar{n}_{\text{H}}} - C_{\text{HII}}(z)\alpha_{\text{B}}(T_{\text{HII}})\bar{n}_{\text{H}}(1+z)^3x_{\text{e}}.
    \label{eq:Q_HII_evo}
\end{equation}
Here we set the escape fraction $f_{\text{esc}} = 0.1$ \citep{Sun-2021}, $\bar{n}_{\text{H}}$ is the mean number density of hydrogen (both neutral and ionized) atoms today, $C_{\text{HII}}=3.0$ is the clumping factor of the ionized gas \citep{Kaurov-2015} which is assumed to be redshift-independent here, $\alpha_{\text{B}}$ is the Case B recombination coefficient, and $T_{\text{HII}}$ is the kinetic temperature. We set $T_{\text{HII}} = 2 \times 10^4$ K as a constant \citep{Robertson-2015}, and then the recombination coefficient $\alpha_{\text{B}} = 2.5 \times 10^{-13}$ cm$^3$s$^{-1}$. 
The total ionization fraction $x_{\text{e}}$ then can be estimated as below, by assuming the helium has the same first stage ionization fraction as hydrogen, and we have
\begin{equation}
    x_{\text{e}} = Q_{\text{HII}} \left( 1 + \frac{Y_{\text{He}}}{4} \right),
    \label{eq:x_e}
\end{equation}
where $Y_{\text{He}} = 0.25$ is the Helium element abundance. The ionizing photons emission rate per unit comoving volume $\dot{n}_{\text{ion}}$ can be estimated by 
\begin{equation} 
    \dot{n}_{\text{ion}} = N_{\text{ion}} \frac{\Omega_{\text{b}}}{\Omega_{\text{m}}} \frac{1}{t_{\text{SF}}(z)} \int_{M_{\text{min}}}^{\infty} dM f_{*}(M) M n(M,z),
    \label{eq:n_ion}
\end{equation}
where $N_{\text{ion}} \approx 4000$ is the total ionizing photons number that a stellar baryon can product throughout its lifetime for typical Pop II galaxies \citep{Leitherer-1999, Vazquez-2005, Leitherer-2010, Leitherer-2014}. $t_{\text{SF}}$ is the star formation time scale, which equals to $10\%$ of the Hubble time at redshift $z$ \citep{Wyithe-2006, Lidz-2011}. 
$M_{\text{min}}$ stands for  the minimum halo mass corresponding to a virial temperature $T_{\text{vir}}$, and halos above this mass can sustain effective cooling via the Ly$\alpha$ transition \citep{Barkana-Loeb-2001}:
\begin{equation}
\begin{split}
        T_{\text{vir}} = 1.98\times10^4 \left( \frac{\mu}{0.6}\right) \left( \frac{M_{\text{min}}}{10^8 h^{-1} M_{\odot}} \right)^{2/3} \\
        \times \left[ \frac{\Omega_{\text{m}}}{\Omega_{\text{m}}(z)} \frac{\Delta_{\text{c}}}{18\pi^2} \right]^{1/3} \left( \frac{1+z}{10} \right) \text{K}, 
        \label{eq:T_vir}
\end{split}
\end{equation}
where $\mu = 0.61$ is the mean molecular weight, $\Omega_{\text{m}}(z)$ is energy density fraction of matter component at redshift $z$, and $\Delta_{c} = 18\pi^2 + 82d - 39d^2$ with $d = \Omega_{\text{m}}(z) - 1$. We taken $T_{\text{vir}} = 10^4$ K to obtain the corresponding $M_{\text{min}}$ at each redshift.

In Eq.~(\ref{eq:n_ion}), we can find that the shape of $f_{*}(M)n(M,z)$ mainly determine the contribution of ionizing photons from galaxies with different masses. In the CDM model, the low-mass galaxies will make a great contribution, but in WDM case, since the formation of low-mass galaxies is suppressed by WDM free streaming effect, the ionizing photons will be dominated by massive galaxies.

In addition, the optical depth of the cosmic microwave background (CMB) scattering is also a good probe for characterizing the reionization history, and it can be estimated by
\begin{equation}
    \tau = \int_{0}^{\infty} \sigma_{\text{T}} \bar{n}_{\text{H}}(1 + z)^3 x_{\text{e}} \frac{cdz}{(1+z)H(z)},
\end{equation}
where $\sigma_{\text{T}} = 6.65 \times 10^{-25}$ cm$^{2}$ is the Thompson scattering cross-section.

\section{Result}\label{sec:result}

\begin{figure*}
    \includegraphics[scale = 0.45]{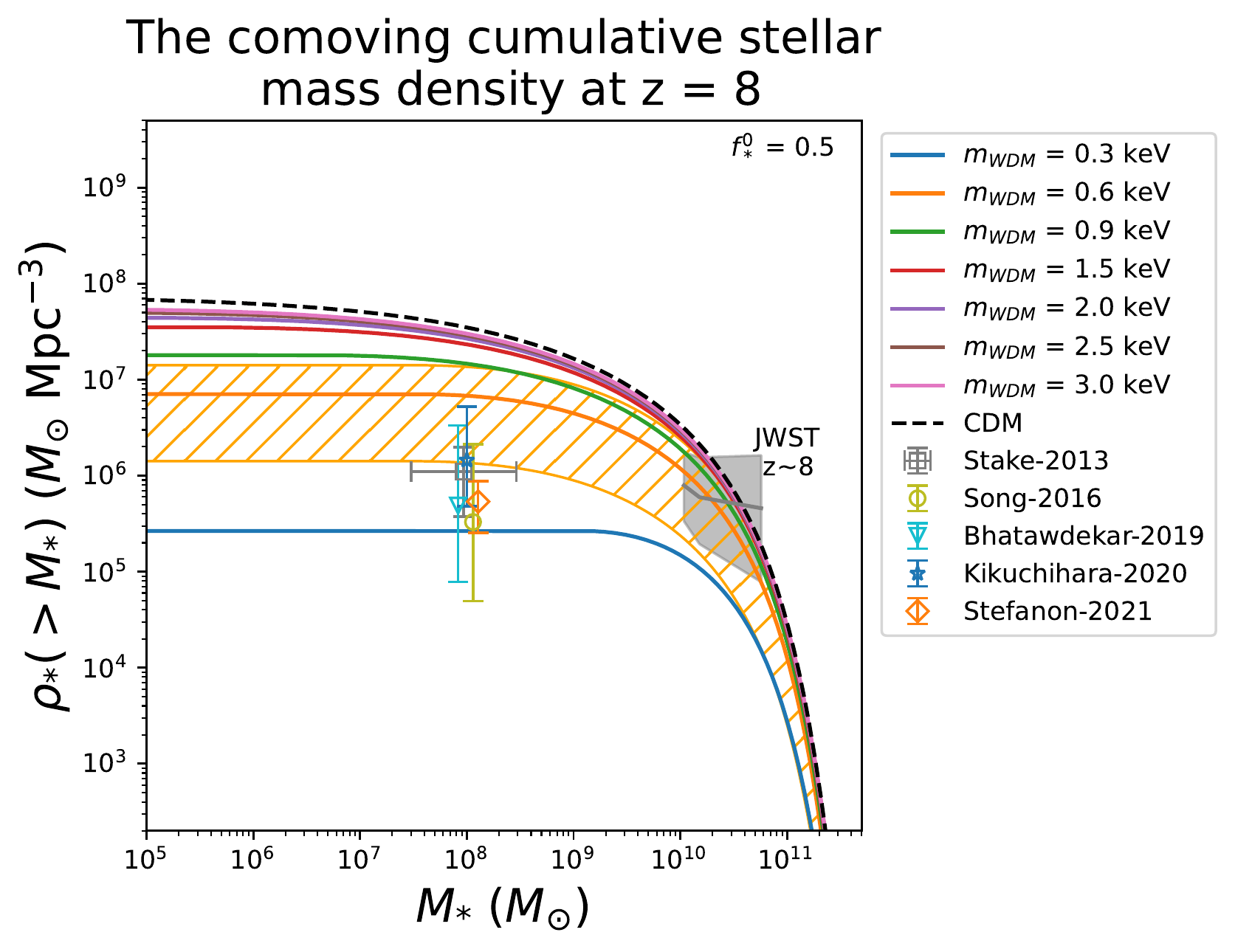}
    \includegraphics[scale = 0.45]{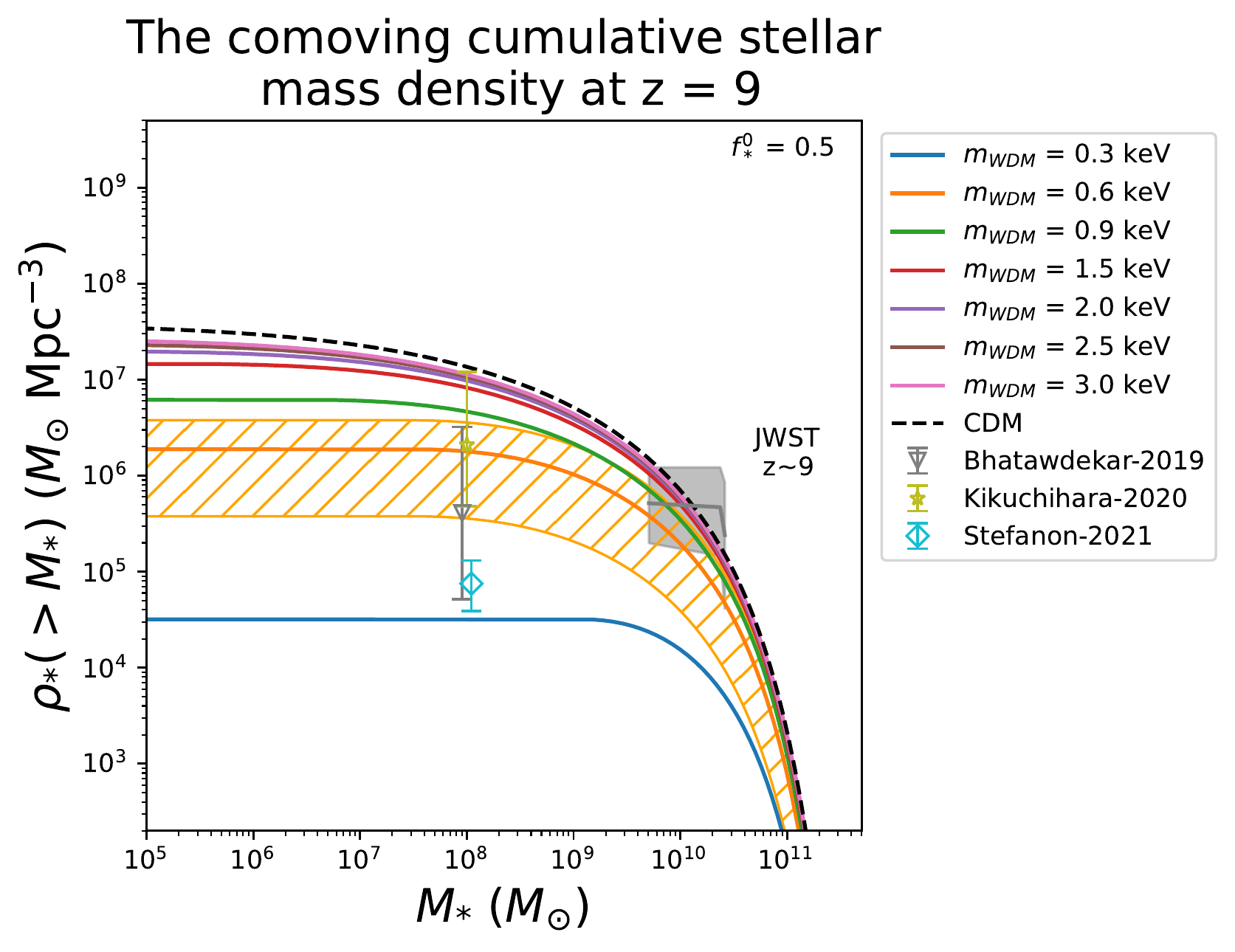}
    \caption{The comoving cumulative stellar mass density for different WDM masses and the CDM case at $z = 8$ (left panel) and $z = 9$ (right panel) with $f_{*}^{0} = 0.5$. We also show the range of the results of $m_{\text{W}} = 0.6$ keV with $f_{*}^{0} = 0.1 - 1.0$ with the orange hatched region. The $\it{JWST}$ results from \cite{Labbe-2023} at $z \sim 8$ and $9$ are shown by the grey shaded regions. We also compare our result with some previous measurements \citep{Stark-2013, Song-2016, Bhatawdekar-2019, Kikuchihara-2020, Stefanon-2021} at low stellar mass region shown as data points with error bars.}
    \label{fig:rho_star}
\end{figure*}

\begin{figure*}
    \includegraphics[scale = 0.4]{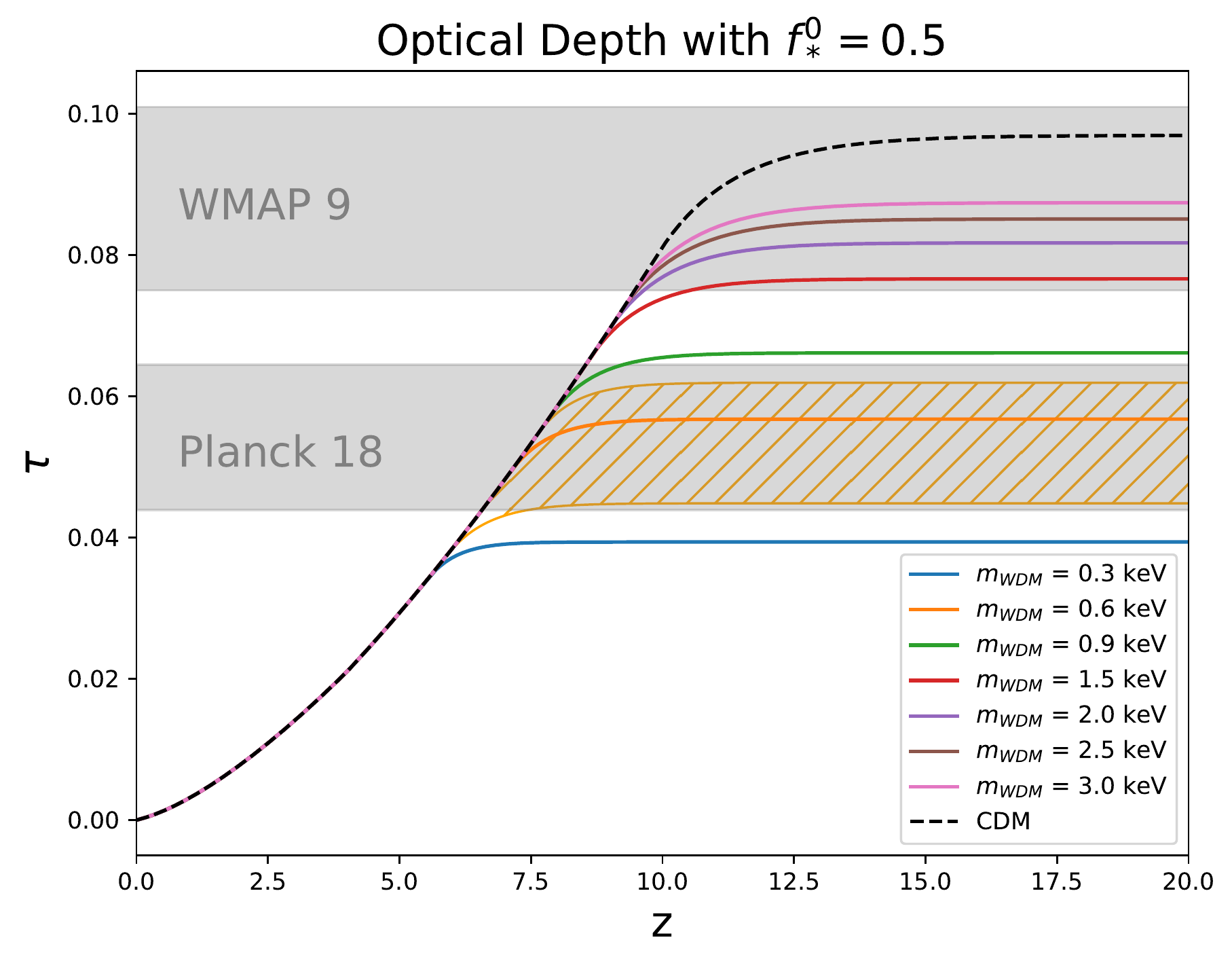}
    \includegraphics[scale = 0.4]{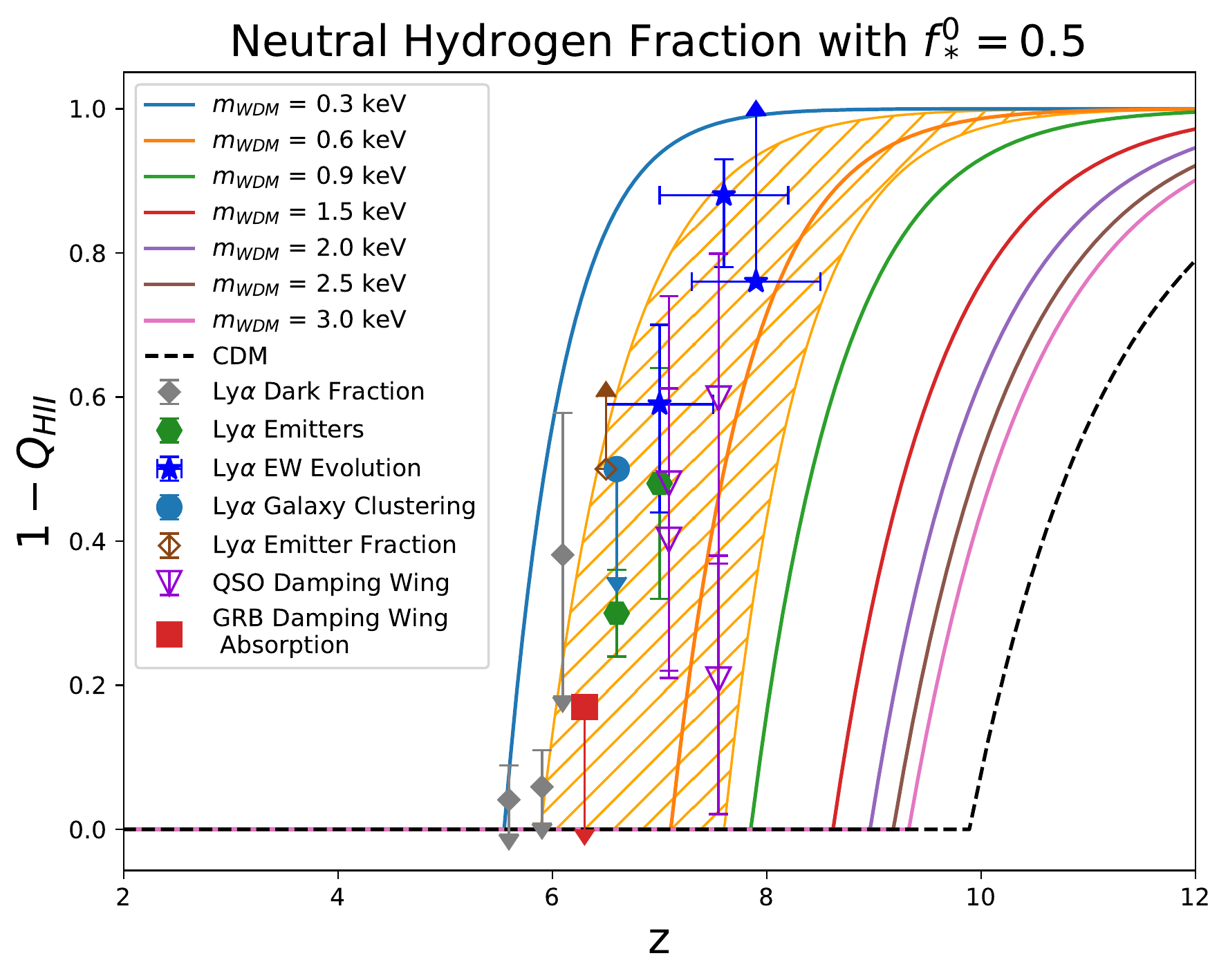}
    \caption{{\it Left panel}: The optical depth $\tau$ as a function of redshift for different WDM masses and the CDM cases with $f_{*}^{0} = 0.5$. The results of $m_{\text{W}} = 0.6$ keV with $f_{*}^{0} = 0.1 - 1.0$ is shown as the orange hatched region. We compare our results with the 1$\sigma$ ($68\%$ C.L.) measurements from $\it{WMAP}$ \citep{WMAP9} and $\it{Planck}$ \citep{Planck18}, which are shown as grey shaded bands. {\it Right panel}: The neutral hydrogen fraction denoted by $1 - Q_{\text{HII}}$ as a function of redshift. The data points are the results from previous observations \citep{Totani-2006, Davies-2018, Greig-2019, Sobacchi-2015, Stark-2011, Dijkstra-2011, Mason-2018, Mason-2019b, Hoag-2019, McGreer-2015, Ota-2008}}
    \label{fig:tau-Q_HII}
\end{figure*}

In Fig.\ref{fig:rho_star}, we present the comoving cumulative stellar mass density at redshift $z = 8$ and $z = 9$ for different WDM masses and CDM case. We find that both WDM and CDM could consistent with $\it{JWST}$ data when the star formation efficiency paramter $f_{*}^{0} \sim 0.5$. But if we compare the result with the previous measurements (e.g. Hubble Ultra-Deep Field, Spitzer/Infrared Array Camera, Keck Observatory, and so on) at low stellar mass region \citep{Stark-2013, Song-2016, Bhatawdekar-2019, Kikuchihara-2020, Stefanon-2021}, we can see that the CDM model will significantly over predict the stellar mass density at this region, since too many low-mass galaxies will form under such high star formation efficiency. On the other hand, the advantage of WDM model is that it will strongly suppress the formation of low-mass galaxies, and is able to match the measurements of both low-mass and high-mass regions. As a reference, we also show the result by varing the star formation efficiency parameter $f_{*}^{0} = 0.1 - 1.0$ with $m_{\text{W}} = 0.6$ keV shown as a hatched orange region.

As we discuss in Sec.\ref{subsec:rho_star-and-reion}, we can also check our result by investigating the impact of our model to the cosmic reionization history. We calculate the optical depth $\tau$ as a function of redshift for the WDM and CDM models, as shown in the left panel of Fig.\ref{fig:tau-Q_HII}, and compare it with the 9-year $\it{WAMP}$ \citep{WMAP9} and $\it{Planck}$ 2018 \citep{Planck18} results. We can see that the model with $m_{\text{W}} = 0.6$ keV is in good agreement with $\it{Planck}$ 2018 measurement.

In the right panel of Fig.\ref{fig:tau-Q_HII}, we present our calculation result of the neutral hydrogen fraction characterized by $1 - Q_{\text{HII}}$ as a function of redshift. The data of observational constraints on the neutral hydrogen fraction from different methods are also shown for comparison, such as Ly$\alpha$ forest dark fraction  \citep{McGreer-2015}, Ly$\alpha$ emitters \citep{Ota-2008}, Ly$\alpha$ equivalent width (EW) distribution \citep{Mason-2018,Mason-2019b,Hoag-2019}, Ly$\alpha$ galaxy clustering \citep{Sobacchi-2015}, Ly$\alpha$ emitter fraction \citep{Stark-2011,Dijkstra-2011}, QSO damping wing \cite{Davies-2018,Greig-2019}, and GRB damping wing absorption \citep{Totani-2006}. We can see that, the WDM model with $m_{\text{W}} = 0.6$ keV and $f_{*}^{0} = 0.1 - 1.0$  (shown by the orange hatched region) is in good agreement with those observational data. So our result indicate that the model with $m_{\text{W}} \sim 0.6$ keV can match both the high redshift comoving cumulative stellar mass density and the reionization history. 

\begin{figure}
    \centering
    \includegraphics[scale = 0.8]{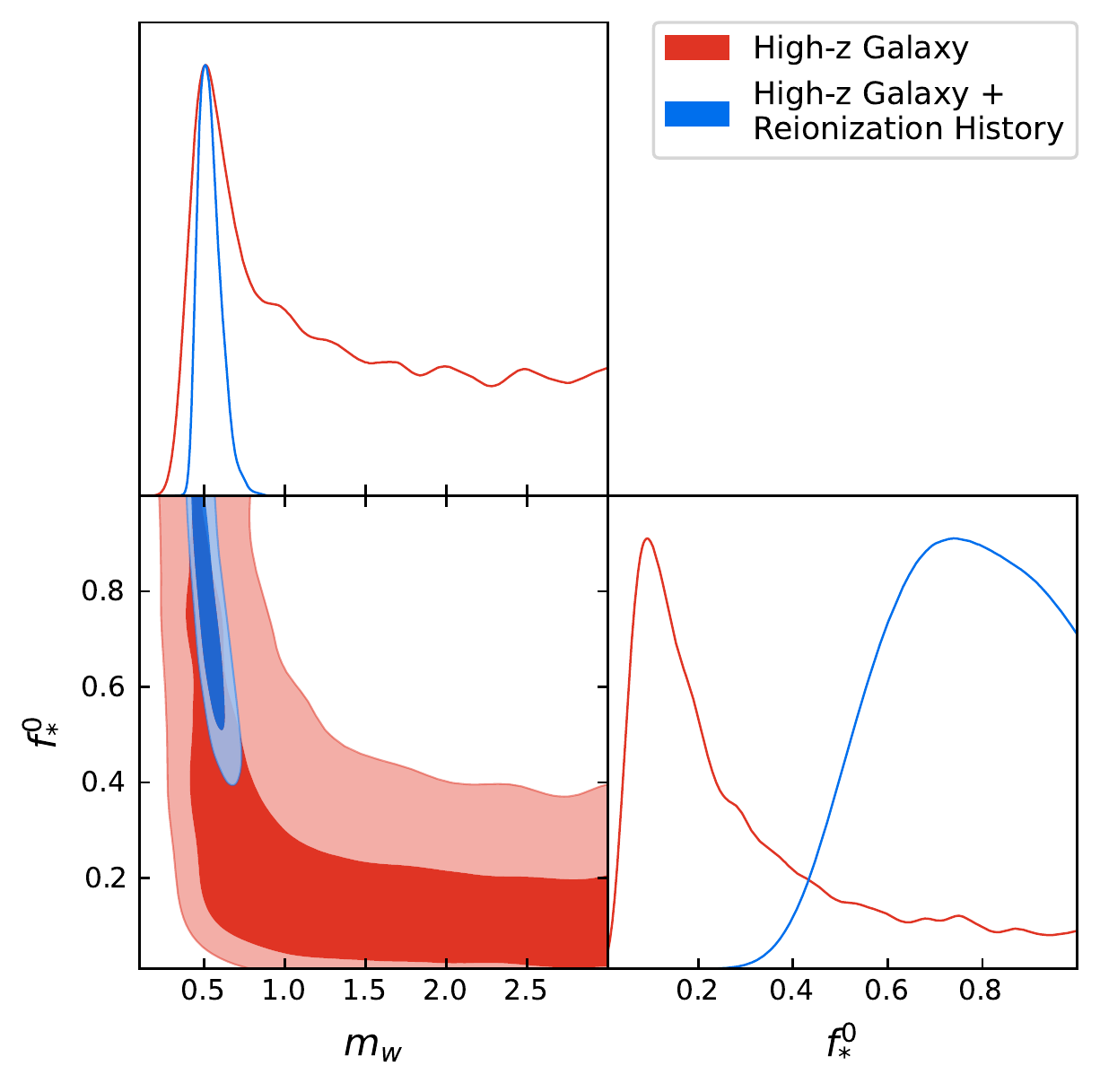}
    \caption{The MCMC constraint result on $m_{\text{W}}$ and $f_{*}^{0}$. Red contours (68.3\% and 95.5\% C.L.) and curves show the constraint from the high-$z$ comoving cumulative stellar mass density data given by {\it JWST} and previous measurements at $z=8$ and 9 shown in Figure~\ref{fig:rho_star}, and the blue contours and curves show the joint constraint result using both the high-$z$ galaxy stellar mass density data and reionization history data (as shown in the left and right panels of Figure~\ref{fig:tau-Q_HII}).}
    \label{fig:MCMC}
\end{figure}

We also sample the posterior distribution of the model parameters using the Markov Chain Monte Carlo method (MCMC) with the {\tt Cobaya} package \citep{cobaya-1, cobaya-2}. The flat priors are used for the two model parameters with $0.1 \leqslant m_{\text{W}}/\text{keV} \leqslant 3.0$ and $0 \leqslant f_{*}^0 \leqslant 1$. Since there is no effective constraint with {\it JWST} data only, we constrain the parameters using the cumulative stellar mass density data from both {\it JWST} and previous measurements at $z=8$ and 9, and also consider the reionization history data, including the optical depth measured by {\it Planck} and the neutral hydrogen fraction measured by various observations (see the data in Figure~\ref{fig:rho_star} and Figure~\ref{fig:tau-Q_HII}). 

We present the constraint result in Fig.\ref{fig:MCMC}. As shown by the red contours, when only considering the cumulative stellar mass density data from {\it JWST} and previous measurements, it favors a large $m_{\rm w}$ and a small $f_{*}^{0}$ which is similar as the CDM case, or a small $m_{\rm w}$ and a large $f_{*}^{0}$. The 1D probablity distribution functions (PDF) of the two parameters are shown in red curves, and we have $m_{\rm w}=0.51^{+1.40}_{-0.10}$ ($1\sigma$) and $0.51^{+2.27}_{-0.17}$ ($2\sigma$) keV and $f_{*}^{0}=0.14^{+0.16}_{-0.07}$ ($1\sigma$) and $0.14^{+0.35}_{-0.10}$ ($2\sigma$). When including the data from reionization history, we find that the constraint on $m_{\rm w}$ is significantly improved with $m_{\rm w}=0.51^{+0.12}_{-0.09}$ ($1\sigma$) and $0.51^{+0.22}_{-0.12}$ ($2\sigma$) keV, and only large $f_{*}^{0}>0.51$ ($1\sigma$) and $>0.39$ ($2\sigma$) is favored in this case.

We notice that, this result may have some tension compared to previous studies. For example, using the UV luminosity functions of high redshift galaxies, a lower limit of $m_{\text{W}} \gtrsim 2.9$ keV can be obtained \citep{Schultz-2014, Menci-2016, Corasaniti-2017, Rudakovskyi-2021}. And a similar result can be derived using the global 21cm signal detected by EDGES (Experiment to Detect the Global Epoch of Reionization Signature) observation \citep{Bowman-2018}, that the WDM models with $m_{\text{W}} \lesssim 3$ keV can be ruled out \citep{Chatterjee-2019}.  We find that, in these works, a low star formation efficiency (e.g. $f_{*}^{0} < 0.1$) is always assumed, which is derived from the observations of Hubble Space Telescope (HST) and other telescopes, and this may be different from the result implied by the $\it JWST$ data given by \cite{Labbe-2023}, which may favor a higher star formation efficiency with $f_{*}^{0} > 0.1$. In Fig.~\ref{fig:dndm}, Fig.~\ref{fig:rho_star}, and Fig.~ \ref{fig:tau-Q_HII}, we plot the results of the WDM model with $m_{\text{W}}=3$ keV for comparison. We can see that the WDM model with $m_{\text{W}}=3$ keV has a suppression mass scale $\sim 10^{7} M_{\odot}$, and obviously cannot fit the data of reionization history with a large star formation efficiency ($f_{*}^{0}=0.5$). A small star formation efficiency with $f_{*}^{0}<0.1$ is needed in this case to match the reionization data, but then the $\it JWST$ data shown in Fig.~\ref{fig:rho_star} cannot be matched, just like the CDM case. Besides, \cite{Enzi-2021} derived a lower limit of $m_{\text{W}} > 6.048$ keV by combining the strong gravitational lensing, the Ly$\alpha$ forest and the abundence of the Milky Way satellites, which depends on the low-mass dark halo mass function and is obviously a challenge to our model. In our low WDM mass case, the formation of the dark halos below $10^{9} \sim 10^{10}$ $M_{\odot}$ are strongly suppressed, and it may be challenging to explain the existence of low-mass galaxies in the Milky Way \citep{McConnachie-2012, Newton-2018}.
Hence, it seems that the inconsistency of the WDM particle mass between our result and the previous studies probably comes from the inconsistency between the data of $\it JWST$ and other previous observations.

We should note that there are indeed some argument on the data given by \cite{Labbe-2023}. For instance, \cite{Prada-2023} found that their galaxy formation model can explains the UV luminosity function measured by $\it{JWST}$ \citep{Naidu-2022-b, Donnan-2023, Harikane-2023} and HST \citep{Oesch-2018, Bouwens-2021, Kauffmann-2022}, and their prediction on star formation rate and stellar mass at $z \sim 8.5$ has good match with the observation of the spectroscopically-confirmed galaxies \citep{Heintz-2023, Fujimoto-2023, Williams-2023, Bouwens-2022}. So they claim that, the stellar mass-to-light ratio during early epochs could not have reached such high values reported by \cite{Labbe-2023}. \cite{Chen-2023} discussed three sources of uncertainties in counting massive galaxies at high-$z$, including cosmic variance, error in stellar mass estimate, and contribution by backsplash. They found that each of them can significantly affect the estimation of stellar mass density at high-$z$. Besides, photometric redshift estimation \citep{Adams-2023} and dust extinction \citep{Ferrara-2022, Ziparo-2023} are also important factors that need to be carefully calibrated in the data analysis. So follow-up measurements by spectroscopic observations (especially by the JWST/NIRSpec) are probably necessary for further confirmation of these data in the future.

\section{Conclusion}\label{sec:conclusion}

Unexpected high stellar mass densities of massive galaxies at $z = 8$ and $9$ have been found by the early $\it JWST$ observations, which may indicate a high star formation efficiency and can have tension with the current measurements of the cosmic reionization history. We propose that the WDM model can be a solution and explore the implications on the properties of WDM particle. Since WDM can suppress the formation of low-mass halos by free streaming effect, the number of small galaxies and hence the ionizing photons will be effectively decreased. Therefore, in this scenario, the reionization history would not be changed with a high star formation efficiency for high-$z$ massive galaxies.

After comparing the predictions of our WDM model to the $\it{JWST}$ (at high stellar masses) and other high-z galaxy observational data (at low stellar masses), the CMB optical depth measurement from $\it{Planck}$, and the neutral hydrogen fraction of IGM measurement from a bunch of different methods, we employ the MCMC method to fit those observational data, and find that the WDM model with $m_{\text{W}} = 0.51^{+0.22}_{-0.12}$ keV with $f_{*}^{0} > 0.39$ is in good agreement with all these observational data in 95.5\% C.L.

We should note that this result still has large uncertainties and may change with the adopted models and observational data. On one hand, there still may be large uncertainties in our current theoretical models and parameters about the WDM, optical depth, and reionization history, which can significantly change the derived WDM particle mass. On the other hand, the early release data of $\it{JWST}$ need to be further confirmed by follow up spectroscopic observations, especially with JWST/NIRCam. If we can obtain higher quality data from future observations of $\it{JWST}$, we are supposed to put a more reliable and tighter constraint on WDM particle mass, and have a better understanding on the nature of dark matter and the galaxy formation process.

\normalem
\begin{acknowledgements}
We acknowledge the support of National Key R\&D Program of China No.2022YFF0503404, 2020SKA0110402, MOST-2018YFE0120800, NSFC-11822305, NSFC-11773031, NSFC-11633004, NSFC-11473044, NSFC-11973047, the CAS Project for Young Scientists in Basic Research (No. YSBR-092), and the Chinese Academy of Science grants QYZDJ-SSW-SLH017, XDB 23040100, and XDA15020200. This work is also supported by the science research grants from the China Manned Space Project with NO.CMS-CSST-2021-B01 and CMS- CSST-2021-A01.
\end{acknowledgements}
  
\bibliographystyle{raa}
\bibliography{bibtex}

\begin{thebibliography}{88}
\providecommand\natexlab[1]{#1}
\providecommand\JournalTitle[1]{#1}

\bibitem[{Adams} {et~al.}(2023)]{Adams-2023}
{Adams}, N.~J., {Conselice}, C.~J., {Ferreira}, L., {et~al.} 2023, \mnras, 518,
  4755

\bibitem[{Barkana} {et~al.}(2001)]{Barkana-2001}
{Barkana}, R., {Haiman}, Z., \& {Ostriker}, J.~P. 2001, \apj, 558, 482

\bibitem[{Barkana} \& {Loeb}(2001)]{Barkana-Loeb-2001}
{Barkana}, R., \& {Loeb}, A. 2001, \physrep, 349, 125

\bibitem[{Benson} {et~al.}(2013)]{Benson-2013}
{Benson}, A.~J., {Farahi}, A., {Cole}, S., {et~al.} 2013, \mnras, 428, 1774

\bibitem[{Bhatawdekar} {et~al.}(2019)]{Bhatawdekar-2019}
{Bhatawdekar}, R., {Conselice}, C.~J., {Margalef-Bentabol}, B., \& {Duncan}, K.
  2019, \mnras, 486, 3805

\bibitem[{Biagetti} {et~al.}(2023)]{Biagetti-2023}
{Biagetti}, M., {Franciolini}, G., \& {Riotto}, A. 2023, \apj, 944, 113

\bibitem[{Blumenthal} {et~al.}(1982)]{Blumenthal-1982}
{Blumenthal}, G.~R., {Pagels}, H., \& {Primack}, J.~R. 1982, \nat, 299, 37

\bibitem[{Bode} {et~al.}(2001)]{Bode-2001}
{Bode}, P., {Ostriker}, J.~P., \& {Turok}, N. 2001, \apj, 556, 93

\bibitem[{Bouwens} {et~al.}(2021)]{Bouwens-2021}
{Bouwens}, R.~J., {Oesch}, P.~A., {Stefanon}, M., {et~al.} 2021, \aj, 162, 47

\bibitem[{Bouwens} {et~al.}(2022)]{Bouwens-2022}
{Bouwens}, R.~J., {Smit}, R., {Schouws}, S., {et~al.} 2022, \apj, 931, 160

\bibitem[{Bowman} {et~al.}(2018)]{Bowman-2018}
{Bowman}, J.~D., {Rogers}, A. E.~E., {Monsalve}, R.~A., {Mozdzen}, T.~J., \&
  {Mahesh}, N. 2018, \nat, 555, 67

\bibitem[{Boylan-Kolchin}(2023)]{Boylan-2023}
{Boylan-Kolchin}, M. 2023, arXiv:2208.01611

\bibitem[{Bullock} \& {Johnston}(2005)]{Bullock-2005}
{Bullock}, J.~S., \& {Johnston}, K.~V. 2005, \apj, 635, 931

\bibitem[{Chatterjee} {et~al.}(2019)]{Chatterjee-2019}
{Chatterjee}, A., {Dayal}, P., {Choudhury}, T.~R., \& {Hutter}, A. 2019,
  \mnras, 487, 3560

\bibitem[{Chen} {et~al.}(2023)]{Chen-2023}
{Chen}, Y., {Mo}, H.~J., \& {Wang}, K. 2023, arXiv e-prints, arXiv:2304.13890

\bibitem[{Cooray} \& {Sheth}(2002)]{Cooray-2002}
{Cooray}, A., \& {Sheth}, R. 2002, \physrep, 372, 1

\bibitem[{Corasaniti} {et~al.}(2017)]{Corasaniti-2017}
{Corasaniti}, P.~S., {Agarwal}, S., {Marsh}, D.~J.~E., \& {Das}, S. 2017, \prd,
  95, 083512

\bibitem[{Davies} {et~al.}(2018)]{Davies-2018}
{Davies}, F.~B., {Hennawi}, J.~F., {Ba{\~n}ados}, E., {et~al.} 2018, \apj, 864,
  142

\bibitem[{de Vega} \& {Sanchez}(2012)]{de-Vega-2012}
{de Vega}, H.~J., \& {Sanchez}, N.~G. 2012, \prd, 85, 043517

\bibitem[{Dijkstra} {et~al.}(2011)]{Dijkstra-2011}
{Dijkstra}, M., {Mesinger}, A., \& {Wyithe}, J. S.~B. 2011, \mnras, 414, 2139

\bibitem[{Donnan} {et~al.}(2023)]{Donnan-2023}
{Donnan}, C.~T., {McLeod}, D.~J., {Dunlop}, J.~S., {et~al.} 2023, \mnras, 518,
  6011

\bibitem[{Endsley} {et~al.}(2022)]{Endsley-2022}
{Endsley}, R., {Stark}, D.~P., {Whitler}, L., {et~al.} 2022, arXiv e-prints,
  arXiv:2208.14999

\bibitem[{Enzi} {et~al.}(2021)]{Enzi-2021}
{Enzi}, W., {Murgia}, R., {Newton}, O., {et~al.} 2021, \mnras, 506, 5848

\bibitem[{Ferrara} {et~al.}(2022)]{Ferrara-2022}
{Ferrara}, A., {Pallottini}, A., \& {Dayal}, P. 2022, arXiv e-prints,
  arXiv:2208.00720

\bibitem[{Finkelstein} {et~al.}(2022)]{Finkelstein-2022}
{Finkelstein}, S.~L., {Bagley}, M.~B., {Haro}, P.~A., {et~al.} 2022, \apjl,
  940, L55

\bibitem[{Finkelstein} {et~al.}(2023)]{Finkelstein-2023}
{Finkelstein}, S.~L., {Bagley}, M.~B., {Ferguson}, H.~C., {et~al.} 2023, \apjl,
  946, L13

\bibitem[{Fujimoto} {et~al.}(2023)]{Fujimoto-2023}
{Fujimoto}, S., {Arrabal Haro}, P., {Dickinson}, M., {et~al.} 2023, arXiv
  e-prints, arXiv:2301.09482

\bibitem[{Gong} {et~al.}(2023)]{Gong-2023}
{Gong}, Y., {Yue}, B., {Cao}, Y., \& {Chen}, X. 2023, \apj, 947, 28

\bibitem[{Greig} {et~al.}(2019)]{Greig-2019}
{Greig}, B., {Mesinger}, A., \& {Ba{\~n}ados}, E. 2019, \mnras, 484, 5094

\bibitem[{Harikane} {et~al.}(2023)]{Harikane-2023}
{Harikane}, Y., {Ouchi}, M., {Oguri}, M., {et~al.} 2023, \apjs, 265, 5

\bibitem[{Heintz} {et~al.}(2023)]{Heintz-2023}
{Heintz}, K.~E., {Gim{\'e}nez-Arteaga}, C., {Fujimoto}, S., {et~al.} 2023,
  \apjl, 944, L30

\bibitem[{Hinshaw} {et~al.}(2013)]{WMAP9}
{Hinshaw}, G., {Larson}, D., {Komatsu}, E., {et~al.} 2013, \apjs, 208, 19

\bibitem[{Hoag} {et~al.}(2019)]{Hoag-2019}
{Hoag}, A., {Brada{\v{c}}}, M., {Huang}, K., {et~al.} 2019, \apj, 878, 12

\bibitem[{H{\"u}tsi} {et~al.}(2023)]{Hutsi-2023}
{H{\"u}tsi}, G., {Raidal}, M., {Urrutia}, J., {Vaskonen}, V., \& {Veerm{\"a}e},
  H. 2023, \prd, 107, 043502

\bibitem[{Jiao} {et~al.}(2023)]{Jiao-2023}
{Jiao}, H., {Brandenberger}, R., \& {Refregier}, A. 2023, arXiv e-prints,
  arXiv:2304.06429

\bibitem[{Kauffmann} {et~al.}(2022)]{Kauffmann-2022}
{Kauffmann}, O.~B., {Ilbert}, O., {Weaver}, J.~R., {et~al.} 2022, \aap, 667,
  A65

\bibitem[{Kaurov} \& {Gnedin}(2015)]{Kaurov-2015}
{Kaurov}, A.~A., \& {Gnedin}, N.~Y. 2015, \apj, 810, 154

\bibitem[{Kazantzidis} {et~al.}(2004)]{Kazantzidis-2004}
{Kazantzidis}, S., {Mayer}, L., {Mastropietro}, C., {et~al.} 2004, \apj, 608,
  663

\bibitem[{Kikuchihara} {et~al.}(2020)]{Kikuchihara-2020}
{Kikuchihara}, S., {Ouchi}, M., {Ono}, Y., {et~al.} 2020, \apj, 893, 60

\bibitem[{Labb{\'e}} {et~al.}(2023)]{Labbe-2023}
{Labb{\'e}}, I., {van Dokkum}, P., {Nelson}, E., {et~al.} 2023, \nat, 616, 266

\bibitem[{Leitherer} {et~al.}(2014)]{Leitherer-2014}
{Leitherer}, C., {Ekstr{\"o}m}, S., {Meynet}, G., {et~al.} 2014, \apjs, 212, 14

\bibitem[{Leitherer} {et~al.}(2010)]{Leitherer-2010}
{Leitherer}, C., {Ortiz Ot{\'a}lvaro}, P.~A., {Bresolin}, F., {et~al.} 2010,
  \apjs, 189, 309

\bibitem[{Leitherer} {et~al.}(1999)]{Leitherer-1999}
{Leitherer}, C., {Schaerer}, D., {Goldader}, J.~D., {et~al.} 1999, \apjs, 123,
  3

\bibitem[{Lewis} {et~al.}(2000)]{CAMB}
{Lewis}, A., {Challinor}, A., \& {Lasenby}, A. 2000, \apj, 538, 473

\bibitem[{Lidz} {et~al.}(2011)]{Lidz-2011}
{Lidz}, A., {Furlanetto}, S.~R., {Oh}, S.~P., {et~al.} 2011, \apj, 741, 70

\bibitem[{Liu} \& {Bromm}(2022)]{Liu-2022}
{Liu}, B., \& {Bromm}, V. 2022, \apjl, 937, L30

\bibitem[{Madau} {et~al.}(1999)]{Madau-1999}
{Madau}, P., {Haardt}, F., \& {Rees}, M.~J. 1999, \apj, 514, 648

\bibitem[{Mason} {et~al.}(2023)]{Mason-2023}
{Mason}, C.~A., {Trenti}, M., \& {Treu}, T. 2023, \mnras, 521, 497

\bibitem[{Mason} {et~al.}(2018)]{Mason-2018}
{Mason}, C.~A., {Treu}, T., {Dijkstra}, M., {et~al.} 2018, \apj, 856, 2

\bibitem[{Mason} {et~al.}(2019)]{Mason-2019b}
{Mason}, C.~A., {Fontana}, A., {Treu}, T., {et~al.} 2019, \mnras, 485, 3947

\bibitem[{McConnachie}(2012)]{McConnachie-2012}
{McConnachie}, A.~W. 2012, \aj, 144, 4

\bibitem[{McGreer} {et~al.}(2015)]{McGreer-2015}
{McGreer}, I.~D., {Mesinger}, A., \& {D'Odorico}, V. 2015, \mnras, 447, 499

\bibitem[{Menci} {et~al.}(2022)]{Menci-2022}
{Menci}, N., {Castellano}, M., {Santini}, P., {et~al.} 2022, \apjl, 938, L5

\bibitem[{Menci} {et~al.}(2016)]{Menci-2016}
{Menci}, N., {Grazian}, A., {Castellano}, M., \& {Sanchez}, N.~G. 2016, \apjl,
  825, L1

\bibitem[{Mirocha} \& {Furlanetto}(2023)]{Mirocha-2023}
{Mirocha}, J., \& {Furlanetto}, S.~R. 2023, \mnras, 519, 843

\bibitem[{Mirocha} {et~al.}(2017)]{Mirocha-2017}
{Mirocha}, J., {Furlanetto}, S.~R., \& {Sun}, G. 2017, \mnras, 464, 1365

\bibitem[{Naidu} {et~al.}(2022{\natexlab{a}})]{Naidu-2022-a}
{Naidu}, R.~P., {Oesch}, P.~A., {Setton}, D.~J., {et~al.} 2022{\natexlab{a}},
  arXiv e-prints, arXiv:2208.02794

\bibitem[{Naidu} {et~al.}(2022{\natexlab{b}})]{Naidu-2022-b}
{Naidu}, R.~P., {Oesch}, P.~A., {van Dokkum}, P., {et~al.} 2022{\natexlab{b}},
  \apjl, 940, L14

\bibitem[{Newton} {et~al.}(2018)]{Newton-2018}
{Newton}, O., {Cautun}, M., {Jenkins}, A., {Frenk}, C.~S., \& {Helly}, J.~C.
  2018, \mnras, 479, 2853

\bibitem[{Oesch} {et~al.}(2018)]{Oesch-2018}
{Oesch}, P.~A., {Bouwens}, R.~J., {Illingworth}, G.~D., {Labb{\'e}}, I., \&
  {Stefanon}, M. 2018, \apj, 855, 105

\bibitem[{Ota} {et~al.}(2008)]{Ota-2008}
{Ota}, K., {Iye}, M., {Kashikawa}, N., {et~al.} 2008, \apj, 677, 12

\bibitem[{Padmanabhan} \& {Loeb}(2023)]{Hamsa-2023}
{Padmanabhan}, H., \& {Loeb}, A. 2023, \apjl, 953, L4

\bibitem[{Parashari} \& {Laha}(2023)]{Prarshari-2023}
{Parashari}, P., \& {Laha}, R. 2023, arXiv:2305.00999

\bibitem[{Pe{\~n}arrubia} {et~al.}(2008)]{Penarrubia-2008}
{Pe{\~n}arrubia}, J., {Navarro}, J.~F., \& {McConnachie}, A.~W. 2008, \apj,
  673, 226

\bibitem[{Planck Collaboration} {et~al.}(2020)]{Planck18}
{Planck Collaboration}, {Aghanim}, N., {Akrami}, Y., {et~al.} 2020, \aap, 641,
  A6

\bibitem[{Prada} {et~al.}(2023)]{Prada-2023}
{Prada}, F., {Behroozi}, P., {Ishiyama}, T., {Klypin}, A., \& {P{\'e}rez}, E.
  2023, arXiv e-prints, arXiv:2304.11911

\bibitem[{Press} \& {Schechter}(1974)]{Press-Schechter}
{Press}, W.~H., \& {Schechter}, P. 1974, \apj, 187, 425

\bibitem[{Robertson} {et~al.}(2015)]{Robertson-2015}
{Robertson}, B.~E., {Ellis}, R.~S., {Furlanetto}, S.~R., \& {Dunlop}, J.~S.
  2015, \apjl, 802, L19

\bibitem[{Rudakovskyi} {et~al.}(2021)]{Rudakovskyi-2021}
{Rudakovskyi}, A., {Mesinger}, A., {Savchenko}, D., \& {Gillet}, N. 2021,
  \mnras, 507, 3046

\bibitem[{Schultz} {et~al.}(2014)]{Schultz-2014}
{Schultz}, C., {O{\~n}orbe}, J., {Abazajian}, K.~N., \& {Bullock}, J.~S. 2014,
  \mnras, 442, 1597

\bibitem[{Sheth} {et~al.}(2001)]{Sheth-Tormen-2001}
{Sheth}, R.~K., {Mo}, H.~J., \& {Tormen}, G. 2001, \mnras, 323, 1

\bibitem[{Sheth} \& {Tormen}(1999)]{Sheth-Tormen-1999}
{Sheth}, R.~K., \& {Tormen}, G. 1999, \mnras, 308, 119

\bibitem[{Sobacchi} \& {Mesinger}(2015)]{Sobacchi-2015}
{Sobacchi}, E., \& {Mesinger}, A. 2015, \mnras, 453, 1843

\bibitem[{Song} {et~al.}(2016)]{Song-2016}
{Song}, M., {Finkelstein}, S.~L., {Ashby}, M. L.~N., {et~al.} 2016, \apj, 825,
  5

\bibitem[{Stark} {et~al.}(2011)]{Stark-2011}
{Stark}, D.~P., {Ellis}, R.~S., \& {Ouchi}, M. 2011, \apjl, 728, L2

\bibitem[{Stark} {et~al.}(2013)]{Stark-2013}
{Stark}, D.~P., {Schenker}, M.~A., {Ellis}, R., {et~al.} 2013, \apj, 763, 129

\bibitem[{Stefanon} {et~al.}(2021)]{Stefanon-2021}
{Stefanon}, M., {Bouwens}, R.~J., {Labb{\'e}}, I., {et~al.} 2021, \apj, 922, 29

\bibitem[{Sun} {et~al.}(2021)]{Sun-2021}
{Sun}, G., {Mirocha}, J., {Mebane}, R.~H., \& {Furlanetto}, S.~R. 2021, \mnras,
  508, 1954

\bibitem[{Torrado} \& {Lewis}(2019)]{cobaya-1}
{Torrado}, J., \& {Lewis}, A. 2019, {Cobaya: Bayesian analysis in cosmology},
  Astrophysics Source Code Library, record ascl:1910.019

\bibitem[{Torrado} \& {Lewis}(2021)]{cobaya-2}
{Torrado}, J., \& {Lewis}, A. 2021, \jcap, 2021, 057

\bibitem[{Totani} {et~al.}(2006)]{Totani-2006}
{Totani}, T., {Kawai}, N., {Kosugi}, G., {et~al.} 2006, \pasj, 58, 485

\bibitem[{Trinca} {et~al.}(2023)]{Trinca-2023}
{Trinca}, A., {Schneider}, R., {Valiante}, R., {et~al.} 2023, arXiv e-prints,
  arXiv:2305.04944

\bibitem[{V{\'a}zquez} \& {Leitherer}(2005)]{Vazquez-2005}
{V{\'a}zquez}, G.~A., \& {Leitherer}, C. 2005, \apj, 621, 695

\bibitem[{Williams} {et~al.}(2023)]{Williams-2023}
{Williams}, H., {Kelly}, P.~L., {Chen}, W., {et~al.} 2023, Science, 380, 416

\bibitem[{Wyithe} \& {Loeb}(2003)]{Wyithe-2003}
{Wyithe}, J. S.~B., \& {Loeb}, A. 2003, \apjl, 588, L69

\bibitem[{Wyithe} \& {Loeb}(2006)]{Wyithe-2006}
{Wyithe}, J. S.~B., \& {Loeb}, A. 2006, \nat, 441, 322

\bibitem[{Yuan} {et~al.}(2023)]{Yuan-2023}
{Yuan}, G.-W., {Lei}, L., {Wang}, Y.-Z., {et~al.} 2023, arXiv e-prints,
  arXiv:2303.09391

\bibitem[{Ziparo} {et~al.}(2023)]{Ziparo-2023}
{Ziparo}, F., {Ferrara}, A., {Sommovigo}, L., \& {Kohandel}, M. 2023, \mnras,
  520, 2445

\end{thebibliography}

\end{document}